\documentclass[letterpaper,aps,twocolumn,groupedaddress,showpacs,amsmath,amssymb]{revtex4-1}
\usepackage[margin=1.55cm]{geometry}
\usepackage{setspace}
\DeclareRobustCommand{\rchi}{{\mathpalette\irchi\relax}}
\newcommand{\irchi}[2]{\raisebox{\depth}{$#1\chi$}}
\usepackage{graphicx}
\usepackage{dcolumn}
\usepackage{bm}
\begin{document}

\title{The magneto-optics in quantum wires comprised of vertically stacked quantum dots: A calling for the
magnetoplasmon qubits}
\author{Manvir S. Kushwaha}
\address
{\centerline {Department of Physics and Astronomy, Rice University, P.O. Box 1892, Houston, TX 77251, USA}}
\hspace{-1.0cm}
\date{\today}

\begin{abstract}
\small{
A deeper sense of advantages over the planar quantum dots and the foreseen applications in the single-electron
devices and quantum computation have given vertically stacked quantum dots (VSQD) a width of interest. Here, we
embark on the collective excitations in a quantum wire made-up of vertically stacked, self-assembled InAs/GaAs
quantum dots in the presence of an applied magnetic field in the symmetric gauge. We compute and illustrate the
influence of an applied magnetic field on the behavior characteristics of the density of states, Fermi energy,
and collective (magnetoplasmon) excitations [obtained within the framework of random-phase approximation (RPA)].
The Fermi energy is observed to oscillate as a function of the Bloch vector. Remarkably, the intersubband
single-particle continuum splits into two with a collective excitation propagating within the gap. This is
attributed to the (orbital) quantum number owing to the applied magnetic field. Strikingly, the alteration in
the well- and barrier-widths can enable us to customize the excitation spectrum in the desired energy range.
These findings demonstrate, for the very first time, the viability and importance of studying the VSQD subjected
to an applied magnetic field. The technological promise that emerges is the route to devices exploiting
magnetoplasmon qubits as the potential option in designing quantum gates for the quantum communication networks.}
\end{abstract}

\maketitle

Notwithstanding that the seeds of their synthesis, fabrication, characterization, and theorization were sown
much before (than recognized) [1-21], it is no secret that the discovery of quantum Hall effects [22-23] had
propelled research on the systems of diminishing dimensions for over two decades [24]. Then suddenly emerged
the experimental extraction of graphene -- a monolayer of sp$^2$-bonded carbon atoms densely packed in a
honeycomb lattice -- in 2004, which motivated researchers to explore other similar materials, such as germanene
and silicene, of the same descent. Promising research in graphene has prompted people to search other layered
materials -- like nitrides, oxides, selenides, sulfides, and tellurides -- to be added to the long list of widely
exploited arsenides and phosphides. The fabrication of nanostructures such as nanoribbons, nanowires, nanoballs,
...etc.
keeps people genuinely fascinated with the miniaturization. However, nothing stands to the ultimate confinement
of the quantum dots, which possess many properties resembling those of the real atoms. The difference is that
many of the electronic, optical, and transport phenomena in the quantum dots are tunable -- unlike in the real
atoms. The tunability offers the prospects of controlling and manipulating the individual charge carriers. This
makes the quantum dots, particularly the self-assembled quantum dots, the real asset for designing nanodevices,
which make the backbone of the future Nanoscience and Nanotechnology.

Given the length scales (of a few nanometers) involved in the experimental setup, we are induced to consider an
infinitely periodic system of two-dimensionally confined (InAs) quantum dot layers separated by GaAs spacers.
This then compels us to recall certain facts which justify the resulting system of quantum wires to be a
potential geometry for the device applications. These are: (i) Sakaki's precise diagnosis, which led him to
propose the designing of such heterostructures with the optical phonon scattering practically eliminated [25];
(ii) the strain due to the lattice mismatch at the interfaces between two (different) hosts is the driving force
for the growth of the self-assembled quantum dots in the final system [26]; (iii) the rigorous justification for
the use of Fermi-liquid-like theories for describing the realistic quantum wires [27]; (iv) the variability of
coupling strength at will not only leads to some significant applications in quantum computation, spintronics,
and quantum communication, but also creates a fertile ground for exploring new fundamental physics such as
manipulating spins and controlling electron g factors [28]; and (v) the strong coupling along the growth direction
offers a quasi-one-dimensional (Q1D) system made up of quasi-zero-dimensional (Q0D) systems and hence fulfills
the yearning for {\em reversing the trend} [24].

Literature bears witness that the very first vertically stacked, self-assembled quantum dots were observed in
InAs islands separated by GaAs spacers along the growth direction in 1995 [26]. Immediately upon being noticed
due to the device potential, this discovery ignited an explosion in investigation into the physics of quantum
dots and a large number of groups became active embarking on a variety of research interests. The result was
a long list of experimental [29-38] and theoretical [39-48] works dealing with elastic, thermal, electrical,
magnetic, electronic, and optical phenomena, which motivated authors to envisage various solid-state devices.
Early efforts had largely focused on the pairs of VSQD separated by thin barriers -- termed quantum dot
molecules (QDM) -- because of their importance in realizing the short-distance quantum-state transfer, which
is essential for the future quantum communication networks. In most applications, it is necessary to stack
multiple quantum dots to allow a larger flux of emitted or absorbed photons. Their number and density depends
on the particular use, however.


In all what has been said and done -- based on the size, shape, and composition -- so far, the focus of almost
all the experimental and theoretical works on the VSQD has largely been limited to a single isolated qubit,
whereas the actual fuss in the quantum communication lies about creating and controlling entanglement of
multiple qubits. Coherent manipulation of exciton qubits seems to be the basic notion in realizing the quantum
gates, which are the building blocks for the quantum computation. The role of an applied magnetic field in
relation with the (nuclear or electron) spin polarization has not been fully appreciated in the system of VSQD.
Despite the acclaimed importance of intra- and inter-subband magnetoplasmon excitations, particularly in the
quantum nanostructures, there is, to the best of our knowledge, no trace of any investigation undertaken with
the intent of offering magnetoplasmon qubits, which can offer a great speed advantage over the exciton qubits.
This letter tends to fill that gap.

\begin{figure}[htbp]
\includegraphics*[width=7cm,height=8cm]{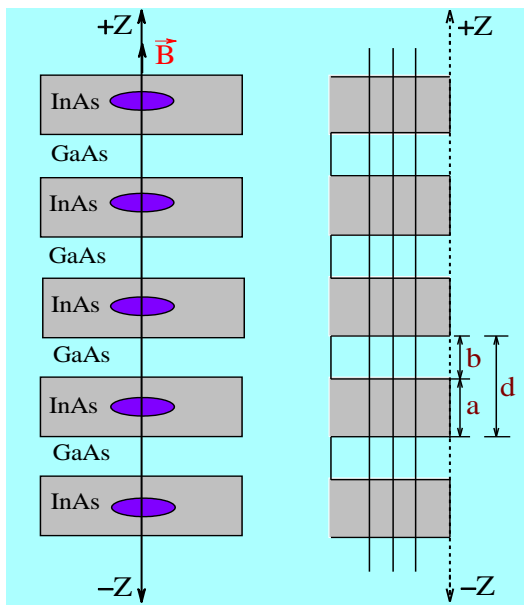}
\caption{Schematics of the quantum wire made up of an infinitely periodic system of InAs islands separated by
GaAs spacer layers (left panel). The right panel shows the Kronig-Penney periodic-potential simulation along
the growth direction. Here $a$ ($b$) is the well (barrier) width and $d=a+b$ is the period of the resultant
system making up the quantum wire.}
\label{fig1}
\end{figure}

We consider a periodic system of quasi-two-dimensional InAs islands of thickness $a$ separated by GaAs spacer layers
of thickness $b$. Each of the InAs island is constrained by a two-dimensional harmonic confining potential of the
form of $V(x)=\frac{1}{2}m^*\omega_o^2 (x^2+y^2)$ in the x-y plane and subjected to an applied magnetic field ($B$)
in the symmetric gauge specified by the vector potential ${\bf A}=\frac{1}{2}\,B$ (-$y$, $x$). The growth (i.e., z)
direction is assumed to be under the greater confinement potential, say, $V_c (z)$, that allows strong coupling
between the InAs islands. The small length scales and strong coupling cause the resultant structure (see Fig. 1) to
mimic a realistic quantum wire with a practically well-defined linear charge density ($n_{1D}$) and justify the
tight-binding approximation (TBA) [40]. The resultant system with moderate tunneling (in the polarizability function)
is describable with the energy dispersion due to tunneling being sinusoidal. Such a system as described above can be
formally characterized by the eigenfunction
\begin{align}
\Psi(r, \theta, z)=\phi_n(r)\,\phi_m(\theta)\,\phi_k(z)\, ,
\end{align}
in the polar coordinates, where
\vspace{-0.25cm}
\begin{align}
\phi_n(r)=\frac{1}{\ell_e}\,\sqrt{\frac{(n+\alpha)!}{n!(\alpha!)^2}}\,e^{-y/2}\,y^{\alpha/2}\,\Phi(-n,\alpha+1,y),\\
\phi_m(\theta)=\frac{1}{\sqrt{2\pi}}\,e^{im\theta}\, ,\\
\phi_k (z)=\frac{1}{\sqrt{N}}\,\sum_l\, e^{i k l d}\,\rchi_t (z - l d)\, ,
\end{align}
where $y=r^2/2\ell_e^2$, $\alpha=\vert m\vert$, $m$ the orbital quantum number, $d=a+b$ the period, $k$ the Bloch
vector, $\ell_e=\sqrt{\hbar/m^*\Omega}$ the effective magnetic length, $\omega_c=eB/(m^* c)$ the cyclotron frequency,  $\Omega=\sqrt{\omega_c^2 + 4\omega_o^2}$ the hybrid frequency of the harmonic oscillator, $\Phi (...)$ the confluent
hypergeometric function, and $\chi_{_{t}}(...)$ the Wannier function; and the eigenenergy
\begin{align}
\epsilon_{nmk}=\!\big[n+\tfrac{1}{2}(\alpha+1)\big]\hbar\Omega\! + \tfrac{1}{2}m\hbar\omega_c\! +
                \epsilon_t\! -\tfrac{W_t}{2}\cos(kd) ,
\end{align}
where $W_t$ is the band-width defined by
\begin{equation}
W_t = - 4 \int^{+a/2}_{-a/2} dz\,\, \rchi_{t} (z)\, V_{_0}\, \rchi_{t}(z-d),
\end{equation}
where we assume the confining potential to be a finite square well (Kronig-Penney potential) with a barrier height
$V_{_0}$ and well-width $a$. Since $N$ (the number of quantum dot layers) is very large, the sum in Eq. (4) can
be written as integral according to the replacement rule: $\sum_k\, \rightarrow \,(N/L_{z})\, \int_{_{BZ}}\,dk$.
Eq. (4) represents the tight-binding constraint which hypothesizes a little overlap between the wave
functions of different sites. Here $\rchi_{t}(...)$, if normalized in the length of the lattice, satisfies:
$\int dz\,\rchi^*_{t}(z-nd)\,\rchi_{t}(z-ld)=\delta_{nl}$ and $\int dz\,\phi^*_{t}(z)\,\phi_{t}(z)=1$. $L_z=Nd$ is
the total crystal length along the growth direction.

For the illustrative examples, we focus on the InAs/GaAs system just as in the original experiment [26]. The material
parameters used are: effective mass $m^*=0.042 m_{_0}$ ($0.067 m_{_0}$) and the background dielectric constant
$\epsilon_{_b}=13.9$ ($12.8$) for the InAs (GaAs). We use the potential barrier of height $V_0=349.11$ meV
that produces the band-width (of the lowest miniband) $W_0=19.76$ meV, in compliance with Sakaki [25] so as to
minimize the optical phonon scattering. The subband spacing $\hbar\omega_{_0}=5.0$ meV, the effective Fermi energy $\epsilon_{eff}=6.4238$ meV for a 1D charge density $n_{1D}=0.7 \times10^6$ cm$^{-1}$, and the effective width of
the confining (parabolic) potential well, estimated as the FWHM of the extent of the eigenfunction,
$w_{eff}=2\sqrt{2 \ln (2)}\sqrt{n+1}\,\ell_{_c}=22.022$ nm. Notice that the Fermi energy $\epsilon_F$ depends upon
the charge density ($n_{1D}$) and the confining potential ($\hbar \omega_{0}$). Thus, we aim at investigating the
magnetoplasmon excitations in a quantum wire made up of VSQDs in a two-subband model within the full RPA [49] at the
absolute zero (T=0 K).

\begin{figure}[htbp]
\includegraphics*[width=7.75cm,height=9cm]{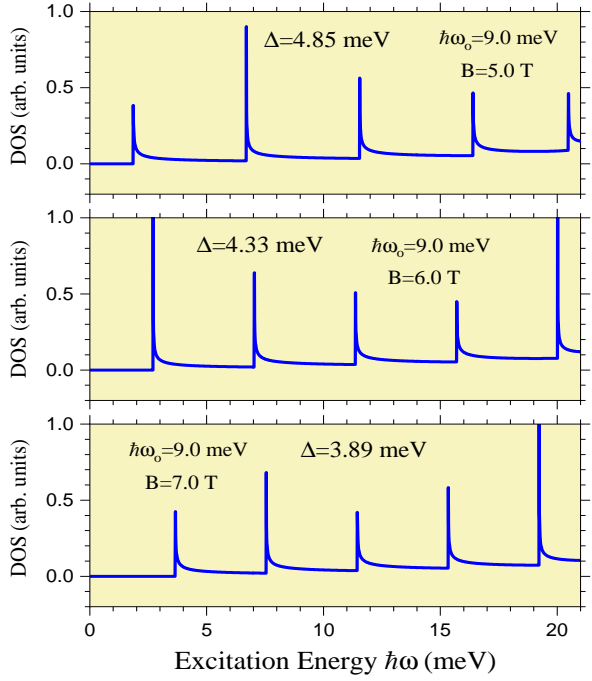}
\caption{The density of states vs. the excitation energy $\hbar\omega$. The confinement potential
$\hbar\omega_o=9.0$ meV and magnetic field $B=5.0$ T (top panel), 6.0 T (middle panel), and 7.0 T
(bottom panel). The band-width is $W_0=19.76$ meV.}
\label{fig2}
\end{figure}

Emergence of the exotic fundamental physics in the past three decades has attested that dimensionality is the most
defining parameter of a material in the condensed matter physics [24]. The density of states (DOS) -- $g(\epsilon)$,
which is essentially the number of different states at a particular energy level that electrons are allowed to
occupy -- is one such preliminary aspect of a system that patently manifests the dimensional dependence. The
analytical results demonstrate that $g(\epsilon) \propto \epsilon^{1/2}$, $g(\epsilon) \propto \epsilon^{0}$, and
$g(\epsilon)\propto \epsilon^{-1/2}$, respectively, in the 3D, 2D, and 1D systems. In the quasi-0D systems --
i.e., the quantum dots -- the DOS is known to be $\delta-$function-like leading to the vanishing of the thermal
broadening [19]. The role of an applied magnetic field is virtually to enhance the confinement for the quantum
dots embedded in the InAs layers. The three panels in Fig. 2 together attest that the DOS peaks start shifting
towards the higher energy and the energy separation between the peaks is reduced with increasing magnetic field.
The realistic quantum wires exhibit similar behavior in the absence of an applied magnetic field, but the DOS
peaks start distancing from each other with switching and increasing the magnetic field [24]. We have noticed that
only the moderate confinement -- with $B\ne0$ -- favors the system of VSQD to impersonate an effective quantum wire.

\begin{figure}[htbp]
\includegraphics*[width=8cm,height=9cm]{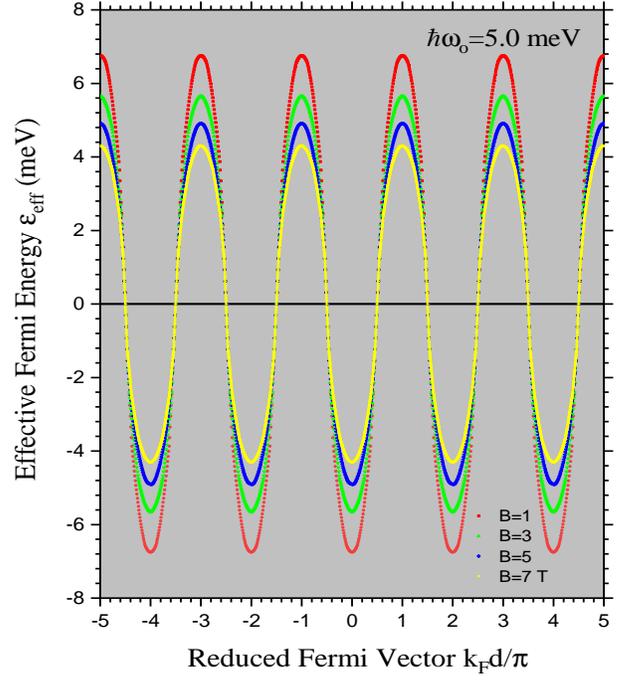}
\caption{The Fermi energy vs. the reduced Fermi vector for the resultant quantum wire, for various values of
the magnetic field $B$. We purposely made this choice to have $k_F d$ on the abscissa -- it could very well
have been the charge density $n_{1D}$. The band-width $W_0=19.76$ meV.}
\label{fig3}
\end{figure}

Figure 3 illustrates the {\em effective} Fermi energy as a function of the reduced Fermi vector ($k_F d/\pi$), for
various values of magnetic field $B$. The Fermi energy is observed to be oscillating as a function of the Fermi
vector $k_F$ for any value of $B$. This oscillatory behavior is not surprising, given the cosine term in the
single-particle energy -- see Eq. (5) -- due to the tight-binding approximation. It is interesting to note
how the peak-to-peak amplitude of the Fermi energy decreases with increasing $B$. This is in compliance with the
fact that the cyclotron radius $r_c \propto 1/B$. Note that the abscissa in Fig. 3 can also constitute the linear
charge density of the system, since the Fermi vector is related to the 1D charge density by a simple relation:
$k_F=(\pi/2) n_{1D}$. The knowledge of the Fermi energy at a given temperature is fundamental to the understanding
of nearly all the electronic, optical, and transport properties of a quantal system. The transport properties (such
as conductance or resistance) are the true reflections of the electron dynamics at the Fermi surface of the system.
What is fascinating about the Fermi surface is that one can tailor it before it tailors the rest in the system.

\begin{figure}[htbp]
\includegraphics*[width=8cm,height=9cm]{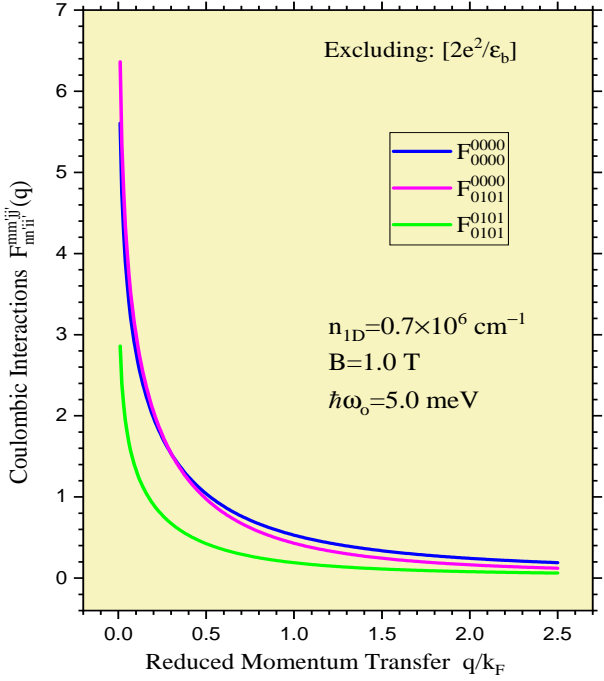}
\caption{(Color online) The Fourier-transformed Coulombic interactions $F^{0000}_{0000}(q)$, $F^{0000}_{0101}(q)$,
and $F^{0101}_{0101}(q)$ plotted as a function of the reduced momentum transfer $q/k_F$. We call attention to the $F^{0000}_{0101}(q)$ being slightly dominant over the $F^{0000}_{0000}(q)$ in the range $0\le q/k_F \le 0.31$ [see
the text].}
\label{fig4}
\end{figure}

Figure 4 exhibits the matrix elements of the Fourier transformed (binary) Coulombic interactions $F^{mm'nn'}_{ii'jj'}$,
where the subscripts (superscripts) refer to the principal quantum number (orbital quantum number). The finiteness of
these elements provides the elementary excitations in the system a multi-particle character by virtue of which the
respective excitations are known as collective excitations: specifically, they are known as plasmons (magnetoplasmons)
depending upon whether the magnetic field $B=0$ ($B\ne0$). The collective excitations are Landau undamped (damped)
before (after) merging with the respective single-particle continuum. We observe that $F^{0000}_{0000}(q) >
F^{0000}_{0101}(q) > F^{0101}_{0101}(q)$ over a large part of the momentum transfer $q$, except for a range of
$0<q/k_F<0.31$ within which $F^{0000}_{0101}(q)$ is slightly greater than $F^{0000}_{0000}(q)$. This specific range is
no trivial thing in the excitation spectrum and can (and, generally, does) have significant consequences. To be brief,
within $0<q/k_F<0.31$, the system gives rise to a metastable state, which is associated with the magnetoplasmon having
a negative group velocity (NGV). The evidence of NGV implies anomalous dispersion in a gain medium with the population
inversion that forms the basis for the lasing action of lasers [50].

\begin{figure}[htbp]
\includegraphics*[width=8cm,height=9cm]{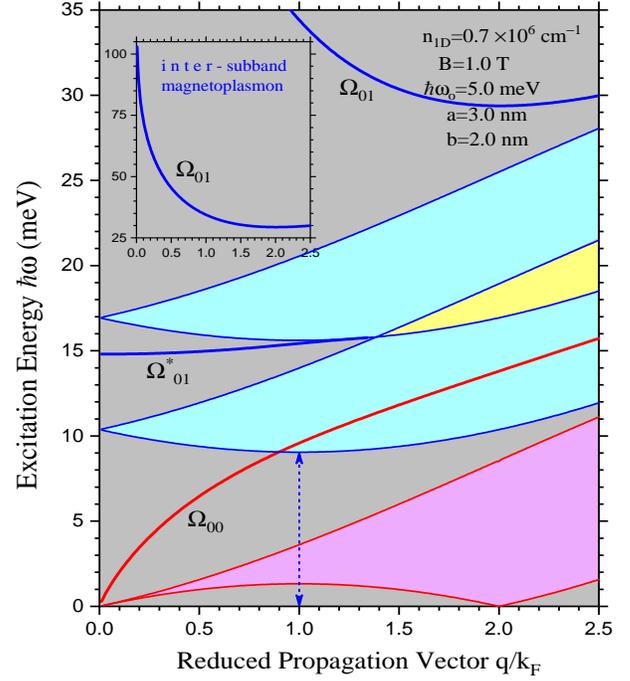}
\caption{The excitation spectrum of the resultant quantum wire within a two-subband model where the excitation energy
$\hbar \omega$ is plotted as a function of the reduced momentum transfer $q/k_F$. The shaded region in magenta (cyan)
refers to the intrasubband (intersubband) SPE associated with the lowest occupied (first excited) subband. The bold
lower (upper) curves in red (blue) represent the intrasubband (intersubband) CME. The intersubband SPE splits into
two with an intersubband CME propagating within the gap. The yellow shade refers to the $\omega$-$q$ space shared by
the split SPE. The vertical double-headed arrow points to the minimum of the intersubband SPE at $q=k_F$. We call
attention to the inset which shows the (upper) intersubband CME (close to zero), which changes the sign of its group
velocity before merging with the upper edge of the respective SPE (at some point in the short wavelength limit). The
parameters used are as listed in the picture.}
\label{fig5}
\end{figure}

Figure 5 portrays the full excitation spectrum of the resultant quantum wire made up of the VSQD in a two-subband
model within the RPA. The plot is rendered in terms of the energy $\hbar\omega$ versus the reduced wave vector
$q/k_F$. The full spectrum consists of the single-particle excitations (SPE) and the collective (magnetoplasmon)
excitations (CME) for a given set of parameters: $n_{1D}$, $\hbar\omega_o$, $B$, $a$, and $b$. The figure caption
specifies the SPE as well as the CME with all the essential details. The intrasubband CME starts from the origin
and propagates without any tendency to merge with the respective SPE up until $q/k_F=2.5$. The intersubband SPE
splits into two with an intersubband CME ($\Omega^*_{01}$) starting from the origin at $\hbar \omega=14.809$ meV
and propagating within the gap to merge with the lower edge of the upper split SPE
(at $q/k_F=1.335$, $\hbar\omega=15.768$ meV), just before the point of intersection of the inner edges of the SPE.
The upper intersubband CME starts from the origin at $\hbar\omega=102.963$ meV with the NGV but changes its sign
at $q/k_F \simeq 2.01$ to propagate with positive group velocity before it merges with the respective SPE in the
short wavelength limit. The most interesting part of the excitation spectrum is that the main intrasubband and
intersubband CME do not merge with the respective SPE, consequently do not suffer from the Landau damping, and
hence remain long-lived excitations for a greater part of the momentum transfer. Such CME make their case quite
easier for the Raman scattering experiments. As to the gap-CME ($\Omega^*_{01}$), its existence is solely
attributed to the applied magnetic field, it is equally bonafide, and should be easily observed in the experiments.
It is not difficult to (analytically) verify the single-particle energies at the critical points: $q/k_F =$ 0, 1,
and 2. Noticeable energy difference between the SPE and the CME at the origin is a manifestation of the many-body
effects such as the depolarization and excitonic shifts [24]. For the set of parameters used in Fig. 5, the average
radius of the quantum dots embedded in the InAs layers is estimated via $1=R^2_o/{2\ell^2_e}$ to be $R_o\simeq19$ nm.


\begin{figure}[htbp]
\includegraphics*[width=8cm,height=9cm]{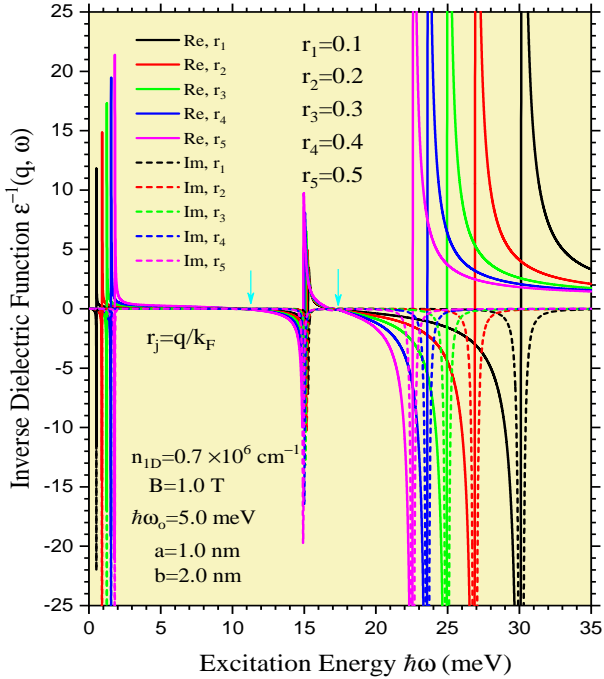}
\caption{(Color online) Inverse dielectric function $\epsilon^{-1}(q,\omega)$ vs. the excitation energy
$\hbar\omega$ for various values of the momentum transfer $q/k_F$. The three peaks, for a given $q/k_F$,
correspond to the three CME in Fig. 5, for example.}
\label{fig6}
\end{figure}

Figure 6 depicts the inverse dielectric function (IDF) versus the excitation energy of the system for various values
of the momentum transfer $q/k_F$. The relevant parameters are as listed in the picture. The real (imaginary) parts
of the IDF are plotted as solid (dashed) curves for a given $q/k_F$. Since we have selected relatively small values
of $q/k_F$, we must expect to reproduce the collective (rather than the single-particle) excitations. It is verified
that all three peaks, for a given $q/k_F$, correspond to the respective CME in the excitation spectrum. A small
energy span covered by the middle peaks -- pertaining to the gap-CME -- indicates that the gap-mode is almost flat
in the long wavelength limit (LWL). The hardcore condensed-matter theorists perceive that searching the poles of the
IDF and the zeros of the DF must, in principle, yield exactly identical results for the excitation spectrum in a
quantal system [51]. However, the former is known to have distinctive advantages over the latter. As an example, the
imaginary (real) part of the IDF provides an appreciable measure of the longitudinal (Hall) resistance in the system.
This suggests that exploring the IDF makes it possible to grasp not only the optical but also the transport phenomena
in the system of interest [52]. Moreover, studying Im[$\epsilon^{-1} (q,\omega)$] also tacitly yields specifics of the
inelastic electron (or light) scattering cross-section in the system. In connection with the inelastic electron
scattering, Im[$\epsilon^{-1}(q, \omega)$] plays a key role in examining such interesting phenomena as the
fast-particle energy loss to a quantum nanostructure. 

In conclusion, we have scrutinized the single-particle and collective (magnetoplasmon) excitations in a quantum wire
comprised of the VSQD in the presence of a planar 2D harmonic confining potential and an applied magnetic field in
the symmetric gauge in a two-subband model within the framework of RPA. The success of RPA in as diverse Q1D systems
as the modulated systems [53] and the Rashba spintronic systems [54] is merely praiseworthy. The notable features of
this investigation are: (i) the $B$-dependence of the DOS substantiating the VSQD mimicking the realistic quantum
wires, (ii) an oscillating Fermi energy as a function of the Fermi vector, (iii) the splitting of the intersubband
SPE at the origin, (iv) the existence of a CME within the gap of the split SPE, (v) all three CME being free from the
the Landau damping up until a relatively large momentum transfer, (vi) the creation of a very high-energy intersubband
CME in the LWL, which starts from the origin and propagates largely with an NGV until a wave vector twice the Fermi
vector, and (vii) the gap-CME propagates to become better defined and tends to remain free from the Landau damping
even in the short wavelength limit ($q\ge 2k_F$) with decreasing period of the system. In a nutshell, the small
length scales and strong coupling along the growth axis effectuate the resultant structure of VSQD to mimic a
realistic quantum wire. This legitimizes our treatment and brings our quest for {\em reversing the trend} to a
fruitful finish. Given the realistic set of parameters used to obtain the results, we are emboldened to believe that
the magnetized chain of VSQD must prove to be a potentially viable system for implementing the formal idea of quantum
state transfer for quantum computing. 





\vspace{0.1cm}
\noindent {\bf Acknowledgment:} The author would like to thank H. Sakaki, P. Nordlander, A.H. MacDonald, and
D. Natelson for the motivating discussions and communication and Kevin Singh for the timely assistance with
the software.




\end{document}